\documentclass[a4paper,11pt]{article}
\usepackage{verbatim}
\usepackage{amsmath}
\usepackage{amsfonts}
\usepackage{amssymb}
\usepackage{multicol}
\usepackage{layouts}
\usepackage{listings}
\usepackage{setspace}
\usepackage{multicol}
\usepackage{titling}
\usepackage{fancyhdr}
\usepackage{graphicx}
\usepackage[usenames]{color}

\usepackage[left=24mm,right=24.5mm,top=45mm,bottom=22mm]{geometry}
\ifx\pdfoutput\@undefined\usepackage[usenames,dvips]{xcolor}
\else\usepackage[usenames,dvipsnames]{xcolor}
\IfFileExists{pdfcolmk.sty}{\usepackage{pdfcolmk}}{}
\fi
\usepackage[plainpages=false,pdfpagelabels,pagebackref=false,naturalnames=true,hyperindex=true,pdftitle={Some Computational Aspects of Essential Properties of Evolution and Life},pdfauthor={Hector Zenil}]{hyperref}
\hypersetup{colorlinks=true,
urlcolor=Cerulean,linkcolor=BrickRed,citecolor=RoyalBlue,a4paper,
 pdfpagemode=None,
 pdfstartview=FitH}
\usepackage[all]{hypcap}

\begin{document}

\title{Some Computational Aspects of Essential\\Properties of Evolution and Life}
\author{Hector Zenil and James A.R. Marshall\\Behavioural and Evolutionary Theory Lab\\Department of Computer Science, The University of Sheffield, UK}

\date{}

\maketitle

\begin{abstract}

While evolution has inspired algorithmic methods of heuristic optimisation, little has been done in the way of using concepts of computation to advance our understanding of salient aspects of biological phenomena. We argue that under reasonable assumptions, interesting conclusions can be drawn that are of relevance to behavioural evolution. We will focus on two important features of life--robustness and fitness--which, we will argue, are related to algorithmic probability and to the thermodynamics of computation, disciplines that may be capable of modelling key features of living organisms, and which can be used in formulating new algorithms of evolutionary computation.\\
 
\end{abstract}

\section{Introduction}

\bigskip

\begin{flushright}
\mbox{
 \begin{minipage}[t][.095\textheight]{.7\textwidth}
\footnotesize It is raining DNA outside. On the bank of the Oxford canal at the bottom of my garden is a large willow tree, and it is pumping downy seeds into the air... It is raining instructions out there; it's raining programs; it's raining tree-growing, fluff-spreading, algorithms. That is not a metaphor, it is the plain truth. It couldn't be any plainer if it were raining floppy disks.\\
\begin{flushright}
Richard Dawkins, 1986\\
\emph{The Blind Watchmaker}\\
Longman Pub
\end{flushright}
 \end{minipage}}
 \end{flushright}

\bigskip
\bigskip
\bigskip
\bigskip
\bigskip
\bigskip

With the work of Darwin it became clear that there were natural processes that shaped features and functions in biological organisms. Evolutionary computation has been inspired by this idea, prompting researchers to posit a correspondence between algorithms and evolution. What evolutionary computation does is to treat natural selection as an algorithmic strategy, iterative in nature, for heuristic problem solving. 

In her contribution to the Ubiquity symposium on the question ``What is computation?" Melanie Mitchell \cite{mitchell} pondered the possibility that biological computation was a process that actually occurred in nature, and that computing would eventually prove as fundamental for biology as physics has been for chemistry.

 Insofar as evolutionary computation models important aspects of evolution, it does so because evolution itself behaves very much like a computational process. If biological evolution is taken to be computational in nature, we can expect it to be driven by the same forces and to be constrained by the same laws that govern information processing.

 Gregory Chaitin, one of the founders of algorithmic information theory \cite{chaitin} (together with A. Kolmogorov \cite{kolmo}, R. Solomonoff \cite{solomonoff} and L. Levin \cite{levin}, among others), has recently suggested \cite{chaitinzenil,chaitinzenil2} that:

\begin{quote}
DNA is essentially a programming language that computes the organism and its functioning; hence the relevance of the theory of computation for biology.
\end{quote}

Indeed, a central element in living systems turns out to be digital: DNA sequences refined by evolution encode the components and drive the development of all living organisms. All examples of life we know have the same (genomic) information-based biology. Information, in living beings, is maintained one-dimensionally through a double-stranded polymer called DNA. Each polymer strand in the DNA contains exactly the same information, coded in the form of a sequence of four different pairs of bases. In attempting to deepen our understanding of life, it is therefore natural to turn to computer science, where the concepts of information, data structures, and algorithms are investigated. 

Important concepts in the theory of computation can help us understand aspects of behaviour and evolution, in particular concepts drawn from algorithmic complexity and computational thermodynamics. Questions such as which genes encode which biological functions, and what shape a protein will fold into, given a DNA sequence, may be effectively formalised in a computational context. We will focus on two pivotal features of life--robustness and fitness--which, we will argue, can be related to properties of computing systems.

%Not too long ago, there was still a pervasive belief that the level of complexity seen in biology was somehow uniquely associated with natural selection. 
%There are many types of search algorithms in existence, of which evolutionary search is an important subset. In computer science, search algorithms define a computational problem in terms of search, where the \emph{search-space} is a space filled with all possible solutions to the problem, and a point in that space defines a solution. 

\section{Biology and Computational Universality}

Evolution manifests some fundamental properties of computation, and not only does it resemble an algorithmic process, it often seems to produce the kind of output a computation would be expected to produce \cite{li}. Witness, for instance, the fact that the instructions for life are stored in sequences of DNA, and in identifiable units of information (genes), even if in a convoluted fashion--full of intricate paths and complicated connections with systematically unpredictable outcomes. 

Computer simulations performed as part of research into artificial life have reproduced various features of evolutionary behaviour \cite{conway,langton,wolfram}, all of which have turned out to be deeply connected to the concept of (Turing) universal computation \cite{turing}. In \cite{margenstern}, for example, an encoding of a very small computer program capable of universal computation is provided as an example of something that, written in the letters of DNA, is smaller than any possible virus. This goes to suggest that it takes very little to attain Turing universality, which is not in any way a special feature of abstract systems. Moreover, it suggests that viruses may potentially be capable of any possible unexpected threat to any possible solution, because they can self-assemble in theoretically any possible configuration (and here we are a level below possible physical constraints), unfamiliar to all previous or current antibodies. Not taking into account this phenomenon of pervasive computational universality in biology, treating it as a mere technicality with little relevance and consequently avoiding it, is a mistake. As we will claim, our knowledge of life may be advanced by studying notions at the edge of decidability and uncomputability (e.g. as we  define it, algorithmic probability is a non-computable measure, but its importance for us lies in the fact that it can be approximated). The concept of Turing universality (and of a universal Turing machine) should simply be treated as a physical system whose richness allows us to study a low level of basic systems behaviour without having to worry about particular causes for particular behaviours. The property that makes a Turing machine Turing universal is its ability to simulate the behaviour of any other computer program or specific Turing machine. Hence its introduction as a tool shouldn't alienate researchers, leading them to treat it as a mere abstract concept with no practical relevance to biology. Some artificial self-assembly models \cite{rothemund} inspired by biological systems, for example, are capable of yielding arbitrary shapes \cite{winfree2} because they turn out to have the property of Turing universality.

 Two of the most important founders of computer science had early interests in biological phenomena. Alan Turing was interested in the question of pattern formation (morphogenesis) \cite{turingmorpho}, while John von Neumann was interested in the problem of self-replication \cite{neumann}, laying the foundations of what is today the theory of cellular automata. The question of self-replication has been generalised to self-assembly \cite{winfree}, and it was von Neumann who first related biological phenomena to (Turing) universality, given that it was this property of computation that ultimately proved that self-replication was not only possible but was indeed a direct consequence of Turing universality. Von Neumann sought to model one of the most basic life processes--reproduction--by designing lattice-based rules where space is updated altogether in discrete steps for studying self-replication. 

Surprisingly, an extremely simple computing system such as Conway's Game of Life \cite{conway} where each cell looks to its neighbours to stay ``alive'' or ``die'' (stay black or white), a two-dimensional cellular automaton, can capture several features that we customarily associate with life (hence its name). At the same time it possesses the most important property of computation, namely Turing universality, thus serving as a strong bridge between these two important scientific disciplines. Today we know that universal systems such as Conway's Game of Life are by no means an exception, often built from a very few, simple components. Other examples have been found (e.g. Langton's ant, Wolfram's Rule 110 and Wolfram's 2, 3 Turing machine). The position that universality is a ubiquitous natural phenomenon has recently been echoed by other respected authors, (see \cite{davis,margenstern}). Wolfram's Principle of Computational Equivalence \cite{wolfram} had in fact already suggested that this was the case.

 Among the key properties of living organisms is their robustness in the face of genetic change \cite{wagner}. Here we propose an indirect connection to Turing universality in order to offer a low level explanation of this biological phenomenon. We think that the properties of algorithmic probability would provide a sound computational framework for explaining biological robustness without leading to any apparent contradiction (such as the one between robustness and evolvability).

\section{Algorithmic Probability and Biological Robustness}

The algorithmic probability \cite{solomonoff,levin} of a string $s$ can be defined as \begin{equation} \label{eq:m} Pr(s) = \Sigma_{p : M(p) = s} 2^{-|p|} \end{equation} That is the sum over all the programs $p$ producing the string $s$. $Pr(s)$ is, therefore, the probability that the output of a (prefix-free\footnote{That is, a machine for which a valid program is never the beginning of any other program, a technicality that allows us to define a probability over strings produced by programs for which the sum is at most 1.}) universal Turing machine $M$, when running a program $p$ comprising a sequence of random (uniformly-distributed) bits before halting, is $s$. 

Levin shows \cite{levin} that there is a universal (semi\footnote{ ``semi'' because, as an uncomputable function, it is approachable from below.}) measure $m$ that dominates any other probability measure $Pr$. Hence we can work with $m$ instead, which thanks to the Levin-Chaitin coding theorem \cite{cover}, can be written as $m(s) \sim 1/2^{K(s)}$, relating the frequency of a pattern to its algorithmic (Kolmogorov-Chaitin) complexity (for a more detailed technical explanation see \cite{algozenil}). 

In order to interpret this computational notion in the context of biology, let's consider a container for each of the four nucleotides of DNA: a (for adenine), c (for cytosine), g (for guanine) and t (for thymine). One could connect the container to a machine, and feed into this ``biological" machine--as input--a random program with instructions constructed from random choices of nucleotides from the containers, forming a sequence of DNA. The machine will then produce an output equivalent to a protein encoded by the ``DNA program'' and will do so precisely according to the distribution of patterns dictated by $m(s)$. Each random program written as a DNA sequence will produce a protein with probability $m(s)$. Of course this is not the way living systems have created DNA sequences, because these sequences are exposed to another, more powerful mechanism: natural selection. But randomness plays an important role in the final construction of these sequences, even if they are selected upon proving to effectively serve a particular use. The instructions that the ribosome follows to compile a protein inside a cell are provided by the mRNA (messenger RNA). This is a polymer formed from the four bases a, c, g, and u (uracil--which replaces thymine (t) during DNA transcription). 

Having learned directly from James D. Watson (the co-discoverer of the structure of DNA), Charles Bennett realised a beautiful analogy which casts RNA polymerase as a specific purpose Turing machine \cite{bennettenergy}, what amounts to a ``truly chemical Turing machine'', an enzyme that crawls along a sequence of instructions analogous to a machine ``tape'' transcribing DNA, stepping left and right just as a Turing machine would (this is not a minor thing; it is the possibility of accessing any part of a Turing machine tape that, alongside other simple factors, contributes to the computational power of a universal Turing machine). The logical state of the RNA polymerase changes according to the chemical information written in the sequence, and Bennett actually used this information to calculate the energy required for this biological process \cite{bennettenergy}. 

Not all biological processes may behave like or necessarily correspond to the exact working mechanism of a Turing machine (in fact this would be an unacceptable assumption), and the concept of computational universality is being introduced only so as to facilitate the consideration of all possible cases (since a machine that can produce any possible sequence and is not constrained to be programmed in any particular way can cover all possible--computable--arrangements). 

It is perhaps not a coincidence that the sequence that the ribosome sequentially ``reads", acting on the directions it provides, is called the ``control tape". This is also very similar to B cells producing a huge variety of antibodies by mixing and matching DNA segments to create antigen-binding sites, picking random combinations of gene segments from many options available from a genome pool \cite{allman}. 

When ribosomes are thought of as computing machines, which is not uncommon in the literature (e.g. see \cite{nature}), this scenario, where $m(s)$ can describe an algorithmic process, is consistent with a information-processing framework, where principles of computation, in particular the distribution of patterns described by $m(s)$ as taking place in the mapping of DNA sequences into proteins (and ultimately into biological functions and shapes), would apply. 

Trying to feed a machine with random sequences of nucleotides will certainly prove to be a frustrating experience. But for those outcomes that are biologically meaningful, algorithmic probability would describe a frequency of distribution of patterns in terms of their algorithmic complexity. Of course, not all programs will produce a biologically viable output. Most will be disappointing, if they produce anything at all, just like Turing machines that may or may not stop, or stop after only one step. These can justifiably be considered to have no relevance to the purpose at hand. Since each amino acid is encoded by three DNA bases, the addition or lack of only one or two bases, for example, would cause a shift in the ``reading process'', making it meaningless. This does not interfere with the claim that $m(s)$ can impact the way the encoding and decoding process takes place in biological systems. Even if one thinks of it as marginal, the fact that one can make any prediction about the distribution of protein folding complexity based on its frequency of occurrence is a reasonable low level explanatory approach to this biological phenomenon, and is consistent with what we find in natural and artificial systems \cite{kamalzenil}, as we will argue in the next section.

 It is important to point out that we are not suggesting that this low level process of structure generation substitutes for any mechanism of natural selection. We are suggesting that computation produces such patterns and natural selection picks those that may provide an evolutionary advantage. If the proportion of patterns (e.g. skin patterns in living systems) are distributed as described by $m(s)$, one may be able to support this basic model with statistical evidence (or disprove it). We have reason to believe that there is evidence that $m(s)$ may be the cause or can describe the distribution of biological patterns (see \cite{zenilmorpho}). 
 
In biology, robustness manifests as the ability of an organism to preserve its phenotype in spite of random mutations or transitional environmental changes. A system is robust if its behaviour remains qualitatively unchanged in the face of sudden random perturbations. The property of being able to produce persistent structures from randomness is essential to our algorithmic model as applied to biology. Start with a random-looking string and run a randomly chosen program on it, and there is a good chance that a high probability random-looking string will be turned into a regular, highly structured one (with low--Kolmogorov--algorithmic complexity). The programs producing these structured strings may be the kinds of building blocks used by organisms and are thought to be common to living systems as basic constituent units \cite{homeobox}. For example, in an experiment with digital organisms designed to show how complex features originate from random mutation and natural selection, it was found that most evolutionary paths build over prior intermediate states to arrive at the same evolved complex functions \cite{lenski}. From the stable frequency of the outputs described by $m(s)$, one can see how distinct complex structures can arise simply from running ``random" programs, quickly amounting to the same solutions. Yet living systems evolve, and are not immune to changes induced by their environments, changes which ultimately favour certain mutations. Chaitin has pointed out \cite{chaitinzenil} something that seems compatible with both the notion of building blocks of life and the notion of biological modularity \cite{wagner2}, viz. that mutations are likely to happen at the level of the program. This means that the system somehow remains consistent at the level of program specification. The mutation is therefore not exactly random at the point level, which in terms of our algorithmic model, remains stable, since most programs produce the same structured output as opposed to a uniform distribution, where any change would yield a noticeable outcome and greater fragility (the assumption behind this argument is that it is the structured output that matters for robustness, and not the random-looking output, which appears more fragile in the distribution of patterns according to $m(s)$). 

For example, one answer to the question of convergent evolution (same solutions to different environments) suggested by this approach is that there are far fewer possible programs producing different outputs and program variations (subroutine changes) than there are possible simple combinatorial solutions, hence reducing the solution space from a possibly uniform distribution to one according to $m(s)$ (a power law distribution) that favours, in frequency and complexity, simple structured outputs. In fact we are investigating this very aspect in connection to a witnessed simplicity bias in natural and artificial genotype-phenotype maps \cite{kamalzenil}.

\section{Computational thermodynamics and Fitness}

On the other hand, if biological systems can be regarded as computational systems one may inquire into biological systems as one would into computational physics. Organisms use information about their environment \cite{valone} to determine the value of different environmental parameters. For example, according to \cite{galef}, learning is a set of complex ontogenetic processes that allows animals to acquire, store, and subsequently use information about the environment. 

To grasp the role of information in biological systems think of a computer as an idealised information-processing system. Today it is fairly easy to understand, from a practical point of view, how energy may be converted into information \cite{landauer,bennettenergy,feynman}. Computers may be thought of as engines transforming energy into information (the information may already be there, but without energy one is unable to extract it). One need only connect one's computer to the electrical grid in order to have it perform a task and produce or reveal information.

 This works in reverse as well, that is, information can be converted into energy, as has been studied in the thermodynamics of computation. A good way to understand the process is by using the well-known thought experiment known as Maxwell's demon paradox \cite{bennettenergy,bennett3}. The basic idea is that one can use one's knowledge about the micro-state of a system to make it $hotter$. Using bits to produce energy is well explained in \cite{feynman} and \cite{sethna}. The connections to information theory and algorithmic complexity, however, are less well understood, but we have a good sense of how we might use algorithmic information theory to connect these concepts at a deep level. 

It is clear that animals convert information into energy--for example, by using information to locate food, to navigate within a habitat, or by learning how to hunt. Similarly, it is clear that energy can be used to extract even greater quantities of energy from the environment--through investment in the construction and operation of a brain. From this it follows that the extent to which an organism is adapted to or is able to produce offspring in a particular environment depends on a positive exchange ratio between information and energy. 

Information processing in organisms should be understood as the process whereby the organism compares its current knowledge about the world with the observable state of the world at a given time. In other words, an animal weighs every possible future outcome against its current representation. While the larger the number and the more accurate the processed observables from the environment the better the decisions and predictions, organisms cannot spend more than the total return received from the use of that information. 

In the context of living organisms, a recent paper \cite{mcnamara} has shown that, under certain assumptions, information processing by individuals can only be a fitness enhancing property. But from computational thermodynamics it follows that any system with access to a finite amount of resources (e.g. memory) has to incur a cost, assuming nothing other than information processing. 

Exchanging information for energy and energy for information some energy inevitably escapes into the environment. Dissipation is a general phenomenon in the real world and it tells us that something is lost in the exchange process, something which itself interacts with the environment, affecting other organisms. In accordance with the second law of thermodynamics, one can see this as information about the system's irreversibility. In biology this kind of exchange happens all the time. Neurons, for example, dissipate about $10^{11}kT$ per discharge \cite{bennettenergy}. Computers (mainly because of their volatile memory devices--the RAM memory) also dissipate energy by at least $1kT$ joules per bit \cite{neumann,landauer,bennettenergy} (where $k$ is the Boltzmann constant and $T$ the ambient temperature), which is the reason computers heat up and require an internal fan. Living cells are today commonly represented in systems biology (even in textbooks, e.g.  \cite{sneppen}) as information processing machines, with E. Coli, given as an example, transcribing $5\times 10^6$ genes during $1/2$ h, i.e. about 10 Gb/h of information. Of course this dissipation is negligible compared to the dissipation of human activity in the transformation and use of energy, but its source is of the same thermodynamic nature; it just belongs to a very different level.

Hence, if an organism aims to gather information about the world, updating its previous state by replacing old information with new (learning), a cost is unavoidable, and takes the form of dissipation. The connection to fitness is then quite natural. Living systems can be regarded as systems with access to finite resources and subject to the same costs as any other physical system ultimately constrained to this rather computational limit.

It is obvious that natural selection has equipped organisms to cope with such trade-offs, to decide whether to undertake certain actions in return for spending resources. This happens, for example, with some animals, who spend most of their energy in the very first instants of a hunt. Because they are not able to keep up the speed of pursuit for long periods of time, they have to ponder and hit upon the optimal starting point for a chase. So information (e.g. cues indicating the location of the prey) can be a fitness enhancing property if the energy return is greater than the cost of processing it and taking action.

Szil\'ard established a connection between energy and information \cite{szilard} at a very low level. The connection is that information can be exchanged for energy because it can help to extract energy. Gaining information lowers the entropy (uncertainty) of a system. The use of a fundamental unit of information to produce energy is well explained in \cite{feynman} and \cite{sethna}. Szil\'ard showed that if one had one bit of information about a system, one could use that information to extract an amount of energy given by $W=kT \log(2)$. Szil\'ard's result in fact implies the 2nd. law of thermodynamics and can be written as follows \cite{broeck}:

\begin{equation}
W_{in} \geq kT \log(2) \geq W_{out} 
\end{equation} 

In other words, one can extract at most $kT \log(2)$ of work from a bit, and it costs at least $kT \log(2)$ to erase (reset) a bit, with $M$ the number of bits to erase (or reset). Landauer studied this thermodynamical argument \cite{landauer} and proposed a principle: if a physical system performs a logically irreversible classical computation, then it must increase the entropy of the environment with an absolute minimum of heat release of $W$ per lost bit. Landauer's principle has recently been tested in the laboratory \cite{landauerexp}.

As pointed out in \cite{mecstats}, natural selection can be seen as extracting information from the environment and coding it into a DNA sequence. But current efforts in the direction of quantifying information content typically do not venture beyond Shannon's communication theory. Nevertheless, computational thermodynamics can help us understand how selection entails the accumulation and exchange of information and energy with the environment in the replacement of populations, as it has already helped to connect computation and physics. 

From the thermodynamics of computation it can be formally concluded, for example, that learning is an energy-saving strategy while ``forgetting'' takes work and consumes information-processing energy (as proven by Landauer and Bennett, this cost is more fundamental, although perhaps more marginal than others). Organisms, therefore, will make an effort to learn fast and keep a structured representation of the world. It also tell us that they wouldn't survive in a random environment or without storing information resources. This means that some predictability is necessary for their survival, and the existence of living systems constitutes a thermodynamical proof of the structure of nature, a position advanced before in, for example \cite{algozenil,zenilinfo}, from the perspective of algorithmic probability (and therefore connecting the two aspects that we are covering in this manuscript). 

We think that the quantification of these dynamics and the trade-offs involved represents an invaluable tool for biological modelling and understanding, and also for evolutionary computation, given that the heuristics of a random search needs to take into account realistic measures for the definition of fitness, fundamental to the description and solution of problems of optimisation.

\section{Concluding Remarks}

%After the discovery of DNA, it was clear that at least an important part of biological systems and evolution was digital and computational in nature, involving code and devices for copying and reusing code, and for mechanically producing new components from specific instructions. 
As a complement to evolutionary computation, and indeed in recompense for what biology has bequeathed to computation in the form of nature-inspired models, we ought to acknowledge that such an exchange from the theory of computation to evolutionary biology is possible, not only because of the existence of strong similarities, but also because computation may explain fundamental aspects of life in the form of abstract models, as in and of itself it may explain pivotal aspects of biological processes and biological functions. And the purpose of abstraction, as Dijkstra would have said, is not to be vague, but to create a semantic level in which one can be absolutely precise, and certainly make process in the formalisation of some of these aspects.
%We have reason to hope, therefore, that this approach that uses the theory of computation and algorithmic information theory will prove fruitful for investigating issues in biology. 

We have yet to see how these concepts may also shape new evolutionary computational algorithms, refining techniques and perhaps improving our grasp and application of them. In this regard it is important to point out that work using algorithmic probability for optimal search algorithms in artificial intelligence has been developed \cite{hutter}, but has not, to the authors' knowledge, been used to complement current approaches to evolutionary computation. So it would be interesting to incorporate this knowledge and trespass the barriers separating the disciplines of computation and biology at a level not attempted before (moving from computation to biology that is, since there has been movement in the other direction, which has greatly benefitted evolutionary algorithms).

\section*{Acknowledgements}

We wish to thank the editors of this Ubiquity Symposium for their invitation, as well as for their comments, which helped improve this presentation. H. Zenil also wishes to thank the Foundational Questions Institute (FQXi) for collaboration support under mini-grant number FQXi-MGA-1212 ``Connections of Computation and Biology", and the Silicon Valley Community Foundation (for mini-grant number 2012-96393 (4661)).

\end{document}